\documentclass[12pt]{article}
\pdfoutput=1
\usepackage{amsmath}
\usepackage{amssymb}
\usepackage{epsfig}
\usepackage{cite}
\usepackage{fontenc}
\usepackage{graphicx}
\usepackage{float}
\usepackage[lofdepth,lotdepth,caption=false]{subfig}
\usepackage{color}
\usepackage{booktabs}
\usepackage{tabularx}
\usepackage{multirow}
\usepackage{longtable}
\usepackage{lscape}
\usepackage{hyperref}
\usepackage{dcolumn}
\usepackage{rotating}

\usepackage[normalem]{ulem}
\definecolor{darkgreen}{cmyk}{1,0,1,0.4}
\definecolor{brown}{cmyk}{0,0.8,1,0.2}
\definecolor{darkred}{cmyk}{0,1,1,0.2}

\allowdisplaybreaks[4] % for page break
\makeatletter
\renewcommand{\fnum@table}{\textbf{\tablename~\thetable}}
\renewcommand{\fnum@figure}{\textbf{\figurename~\thefigure}}
\makeatother

\newcounter{myenumi}

\renewcommand{\themyenumi}{\roman{myenumi}}
{\end{list}}

\newlength{\myem}
\newcounter{mysubequation}[equation]

\makeatletter
\renewcommand{\section}{\@startsection{section}{1}{0em}{-\baselineskip}%
{\baselineskip}{\normalfont\large\bfseries}}
\renewcommand{\subsection}%
{\@startsection{subsection}{2}{0em}{-0.7\baselineskip}%
{0.7\baselineskip}{\normalfont\bfseries}}
\makeatother

\newcommand{\bi}{\begin{itemize}}
\newcommand{\ei}{\end{itemize}}

\newcommand{\bea}{\begin{eqnarray}}
\newcommand{\eea}{\end{eqnarray}}

\newcommand{\nue}{\ensuremath{\nu_e}}
\newcommand{\numu}{\ensuremath{\nu_\mu}}

\def\epsilon{\varepsilon}

\newcommand\schd{Schr$\ddot{\rm o}$dinger}

\def\ket#1{| \,#1\, \rangle}

\def\<{\langle}
\def\>{\rangle}

\def\dfrac#1#2{{\displaystyle\frac{#1}{#2}}}

\def\lsim{\mathrel{\rlap{\lower4pt\hbox{\hskip1pt$\sim$}}
    \raise1pt\hbox{$<$}}}         %less than or approx. symbol
\def\gsim{\mathrel{\rlap{\lower4pt\hbox{\hskip1pt$\sim$}}
    \raise1pt\hbox{$>$}}}         %greater than or approx. symbol

\newcommand{\pab}[1]{\ensuremath{{ P}_{ab}}}
\newcommand{\pba}[1]{\ensuremath{{ P}_{ba}}}
\newcommand{\pbarab}[1]{\ensuremath{{ P}_{\bar{a} \bar{b}} }}
\newcommand{\pbarba}[1]{\ensuremath{{ P}_{\bar{b} \bar{a}} }}
\newcommand{\acpab}[1]{\ensuremath{A^{CP}_{a b}}}
\newcommand{\acpaa}[1]{\ensuremath{A^{CP}_{a a}}}
\newcommand{\acpba}[1]{\ensuremath{A^{CP}_{b a}}}
\newcommand{\atab}[1]{\ensuremath{{A}^{T}_{a b}}}
\newcommand{\acptab}[1]{\ensuremath{A^{CPT}_{a b}}}
\newcommand{\dpcpab}[1]{\ensuremath{\Delta {\cal P}^{CP}_{a b} }}
\newcommand{\dptab}[1]{\ensuremath{\Delta {\cal P}^{T}_{a b}}}
\newcommand{\dpcptab}[1]{\ensuremath{\Delta {\cal P}^{CPT}_{a b} }}

\begin{document}
%
%%%%%%%%%%%%%%%%%%%%%%%%%%%%%%%%%%%%%%%%%%%%%%%%%%%%%%%%%%%%%%%%%%%%%
%%%%                     Title-page                              %%%%
%%%%%%%%%%%%%%%%%%%%%%%%%%%%%%%%%%%%%%%%%%%%%%%%%%%%%%%%%%%%%%%%%%%%%

\begin{titlepage}

\renewcommand{\thefootnote}{\alph{footnote}}

\vspace*{-3.cm}
\begin{flushright}
% Report numbers

\end{flushright}

\renewcommand{\thefootnote}{\fnsymbol{footnote}}
\setcounter{footnote}{-1}

{\begin{center}
{\large\bf  Impact of Matter Density Profile Shape on NSI at DUNE  
\\[0.2cm]
}
\end{center}}

\renewcommand{\thefootnote}{\alph{footnote}}

\vspace*{.8cm}
\vspace*{.3cm}
{\begin{center} 
            {{\sf 
                Animesh Chatterjee $^\star$~\footnote[1]{\makebox[1.cm]{Email:}
                animesh.chatterjee@uta.edu},
				Felipe Kamiya$^\dagger$~\footnote[2]{\makebox[1.cm]{Email:}
                felipe.kamiya@ufabc.edu.br}, 
                Celio A. Moura$^{\dagger}$~\footnote[3]{\makebox[1.cm]{Email:}
                celio.moura@ufabc.edu.br} and
                Jaehoon Yu $^\star$~\footnote[4]{\makebox[1.cm]{Email:}
                jaehoon@uta.edu}
               }}

\end{center}}
\vspace*{0cm}
{\it 
\begin{center}
%\footnotemark[1]%
$^\star$\, Department of Physics, University of Texas at Arlington, Arlington, TX 76019, USA 

%\footnotemark[2]%
$^\dagger$\, Centro de Ci\^encias Naturais e Humanas, Universidade Federal do ABC (UFABC), Santo Andr\'e, SP 09210-580, Brazil
		
\end{center}}

\vspace*{1.5cm}

{\Large 
\bf
\begin{center} Abstract 
\end{center} 
 }

We discuss the impact of  matter density profile shape on the determination of non-standard neutrino matter interactions (NSI) in the context of the long baseline accelerator experiments such as Deep Underground Neutrino Experiment (DUNE). The primary scientific goals of DUNE are to determine the neutrino mass hierarchy, the leptonic CP violation phase, and the existence of new physics beyond the standard model of particles. Here we study the role of different earth matter density profiles on the question of observing standard oscillation as wells as NSI at DUNE. We consider two different earth matter density profiles which are relevant for the DUNE baseline. We first discuss the impact of matter on both appearance and disappearance oscillation channels, then we demonstrate the effect of different matter density profiles on the determination of NSI. We consider four different scenarios of NSI and elucidate the effect at the oscillation probability and measurement of number of events at DUNE. In one case of study we show that a non-standard complex phase $\phi_{e\tau}$ could significantly increase the sensitivity to different matter distributions along the baseline.
\vspace*{.5cm}

\end{titlepage}

\newpage

\renewcommand{\thefootnote}{\arabic{footnote}}
\setcounter{footnote}{0}

%----------------------------------------------------------------------%
\section{Introduction}
Neutrino physics offers great potential for revealing the physics beyond the standard model (BSM). Neutrino oscillation and, consequently neutrino mass~\cite{Schechter:1980gr,Altarelli:2004za,Mohapatra:1980yp,Zee:1980ai,Hirsch:2004he}, remains one of the first and few solid empirical indicators of Beyond Standard Model (BSM) physics. The neutrino mixing parameters ($\theta_{12}, \theta_{13},\theta_{23}, \delta{m^{2}_{21}}, \delta{m^{2}_{31}}$)~\cite{Esteban:2016qun,deSalas:2017kay} have been already measured to good precision and this precision is expected to improve as ongoing and near-future experiments reach higher levels of accuracy and statistics. However, there are still open challenges and unanswered questions, namely the Neutrino mass ordering (mass hierarchy) and the exact value, i.e. octant, of the mixing angle $\theta_{23}$, the discovery of leptonic sector CP violation (through the phase $\delta_{\rm cp}$), the existence of sterile neutrino, which must be resolved by neutrino oscillation experiments.\\

A major goal of present and future long-baseline neutrino oscillation experiment is to make precision measurements of neutrino flavor oscillations. The Deep Underground Neutrino Experiment (DUNE)~\cite{Acciarri:2015uup} is a leading-edge, international experiment for neutrino science and proton decay studies. The experimental goals of DUNE are similar to the ones just cited, including the measurement of the neutrino mass ordering, the octant of the atmospheric mixing angle, and whether there is CP violation in the lepton sector.\\ 

The scientific potential of DUNE in the presence of standard oscillations has been studied extensively by the DUNE collaboration~\cite{Acciarri:2016ooe} and others. However, other mechanisms could be responsible for neutrino flavor change on a sub-leading level~\cite{Miranda:2016ptb,Dey:2018yht}. Any sizable new physics effect is expected to modify the event spectrum at the DUNE detector, and hence, its reach~\cite{Deepthi:2017gxg,Deepthi:2016erc,Blennow:2016etl,Bakhti:2016gic,Escrihuela:2016ube}.\\

Once we invoke new physics, it seems rather unnatural to exclude the possibility of non-standard interactions (NSI)~\cite{Escrihuela:2011cf,Miranda:2015dra,Farzan:2017xzy,Denton:2018xmq}, which can, in principle, allow for flavor changing interactions. It has been established that the presence of matter NSI in general reduces the sensitivity of DUNE to standard oscillation parameters~\cite{Chatterjee:2014gxa,Coloma:2015kiu,deGouvea:2015ndi}. The main reason behind this reduction is the interplay between oscillations due to standard and non-standard parameters that gives rise to a few degeneracies in the sensitivity for DUNE~\cite{Masud:2015xva,Masud:2016bvp}. It has been shown~\cite{Coloma:2016gei} that for sufficiently large values of the NSI parameters one could expect a degeneracy between the sign of $\Delta{m^{2}_{31}}$, $\delta_{\rm cp}$, and the measurement of $\theta_{23}$ octant. As the NSI paradigm brings in a large number of parameters, the statistical analysis of the projected data at DUNE becomes cumbersome and challenging.\\

Matter effects on the neutrino oscillations was studied in different contexts as, for instance, in~\cite{Jacobsson:2001zk,Gandhi:2004md,Akhmedov:2004ny}. The neutrino properties get modified due to the effect of matter. Even a massless neutrino acquires an effective mass and an effective
potential in matter~\cite{Valle:1987gv}. For being a long baseline neutrino experiment, matter effect has a crucial role in DUNE physics reach. The impacts have been studied for a long time~\cite{Huber:2004ka} and are critical for the physics goals of the DUNE~\cite{Acciarri:2015uup}.
\\

In this paper, we discuss the impact of different Earth matter density profiles on the measurement of non-standard neutrino interaction at DUNE. We show that, although NSI effect depends on the matter density, the difference between different profiles is not significant for DUNE at leading order, but may be significant for NSI measurements. We also show a comparison of the uncertainty in the event rate with the difference of the event rate for two specific profiles with NSI.
\\

The paper is organized as follows: in section 2, we discuss effects of earth matter density on the neutrino oscillation probability. In section 3, we discuss the effects of NSI on the neutrino oscillation and show the impact of matter density profile. We show the event rates for DUNE detector on section 4. Finally discussion and conclusion on section 5.
%----------------------------------------------------------------------%
\section{Earth matter effects on neutrino oscillation}
Neutrino oscillation arises from a mixture between the flavor and mass eigenstates of neutrinos. The neutrino flavour eigenstate $\ket{\nu_{\alpha}}$($\alpha = e, \mu, \tau$) can be written as a superposition of mass $m_j$ eigenstates $\ket{\nu_{j}}$( j=1,2,3) as
\begin{equation}
\ket{\nu_{\alpha}} = \sum_{j=1,2,3} U^{*}_{{\alpha}j}\ket{\nu_{j}} \,,
\label{eq1}
\end{equation} 
where $U_{{\alpha}j}$ is a 3$\times$3 unitary mixing matrix, known as the Pontecorvo-Maki-Nakagawa-Sakata(PMNS) matrix \cite{pontecorvo,mns}.\\

The oscillation probability from a flavor state $\nu_{\alpha}$ to $\nu_{\beta}$ traveling a distance L can be written as 
\begin{equation}
P_{\nu_{\alpha} \rightarrow \nu_{\beta}}(L) = \sum_{j,k} U^{*}_{{\alpha}j}U_{{\beta}j}U_{{\alpha}k}U^{*}_{{\beta}k}e^{-i(H_{ij})L} \,,
\label{eq7}
\end{equation}
where $H_{ij}$ is the Hamiltonian in the eigenbasis. In Vacuum $H_{ij} = diag(0, \Delta_{21}, \Delta_{31})$, where $\Delta_{ij} = \frac{1}{2E}\delta m^{2}_{ij}$ and $\delta m^{2}_{ij} = m_i^2 - m_j^2$.\\

When neutrinos travel through a dense medium, their propagation can be significantly modified by the coherent forward scattering from particles they encounter along the way. Potential due to the scattering on matter modifies the mixing of the neutrinos. As a result, the oscillation probability differs from the oscillation in vacuum. The effective Hamiltonian in matter can then be written as\\
\begin{equation}
H_{eff} = H_{vac} + H_{mat} \,,
\end{equation}
where $H_{vac}$ is the vacuum Hamiltonian and the matter Hamiltonian for an electron neutrino propagating through matter (electrons) can be written as
\begin{equation}
H_{mat} = \frac{G_{F}}{\sqrt 2}[\bar{\nu_{e}}\gamma^{\mu}(1-\gamma^{5})e][\bar{e}\gamma_{\mu}(1-\gamma^{5})\nu_{e}] \,,
\label{eq20}
\end{equation}
where $G_F$ is the Fermi coupling constant, $\gamma^{\mu,5}$ are Dirac gamma matrices, $e$ and $\nu_e$ are electron and electron neutrino spinors.
If we consider the effective Hamiltonian over the electron background and integrate over the electron momentum,  the Hamiltonian can then be written as
\begin{equation}
H_{eff} = V_{CC} \bar{\nu_{e}} \gamma^{0} \nu_{e} \,,
\label{eq22}
\end{equation}
where the charged current potential 
\begin{equation}
V_{CC} = \sqrt{2}G_{F} N_{e} \,.
\label{eq23}
\end{equation}
Here $N_{e}$ is the electron density of the medium. Similarly, the neutral current potential of neutrinos propagating in a medium will be 
\begin{equation}
V_{NC} = \frac{G_{F}N_{n}}{\sqrt{2}} \,,
\label{eq24}
\end{equation} 
where $N_n$ is the neutron density. It is important to note that the neutral current potential is flavor independent, hence it will not have any effect on neutrino oscillations.\\

It is useful to write the matter potential in terms of the matter density $\rho$ and the electron-fraction in the nucleon $Y_{e}$ as
\begin{equation}
\frac{V_{CC}}{[{\rm eV}]} = 7.56\times 10^{-14}(\frac{\rho}{{\rm [g/cm}^{3}]}) Y_{e} \,. 
\label{eq25}
\end{equation}
The flavor-changing mechanism in matter was formulated by Mikhaev, Smirnov and Wolfenstein (MSW)~\cite{Wolfenstein:1977ue,Mikheev:1987qk}, who first pointed out that there is an interplay between flavor-non-changing neutrino-matter interactions and neutrino mass and mixing. The MSW effect stems from the fact that electron neutrinos (and antineutrinos) have different interactions with matter compared to neutrinos of other flavors. In particular, $\nu_{e}$ can have both charged current and neutral current elastic scattering with electrons, while $\nu_{\mu}$ or $\nu_{\tau}$ have only neutral current interactions with electron. Hence, for long baseline neutrino experiment like DUNE, which focus on the $\nu_{e}$ appearance channel, it is crucial to understand the effect of Earth matter density which changes oscillation probability significantly.

\subsection{Matter density profile effects in the probability of oscillation} 
\begin{figure}[htbp]
\centering
\includegraphics[width=0.8\textwidth]{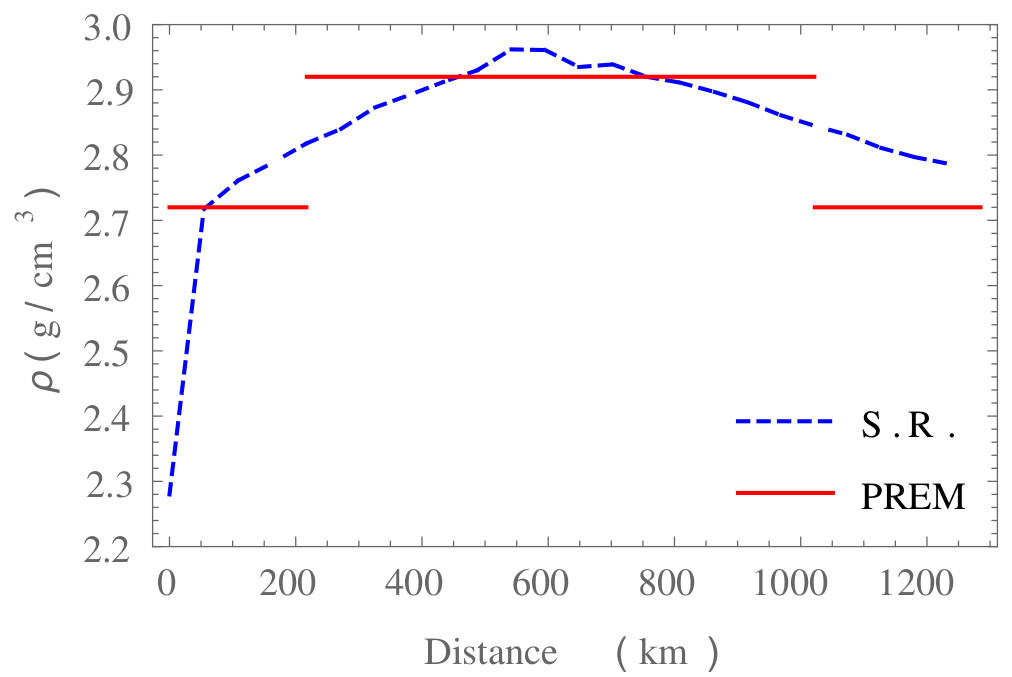}
\caption{Density profiles used for comparison of the NSI effect on neutrino propagation.} \label{fig:profile}
\end{figure}

In the context of the DUNE experiment, the effects of matter density variation and its average along the beam path from Fermilab to SURF were studied considering the standard neutrino oscillation framework with three flavors~\cite{Roe:2017zdw,Kelly:2018kmb}. The DUNE collaboration uses the so called PREM~\cite{PREM,PREM2} density profile to consider matter effects. With this assumption, the neutrino beam crosses a few constant density layers. However, a more detailed density map is available for the USA with more than 50 layers and $0.25 \times 0.25$ degree cells of latitude and longitude: The Shen-Ritzwoller profile~\cite{SR:2016,Roe:2017zdw}. Figure~\ref{fig:profile} shows the two profiles for comparison.

The parameters to calculate the probability of oscillation with their respective true values for normal and inverted hierarchy, NH and IH respectively, are shown in table~\ref{valoresatuais}.
\begin{table}[htb]
\caption{Parameters and considered true values, compatible with~\cite{Gonzalez-Garcia:2013usa}, for normal and inverted hierarchy, NH and IH respectively.} \label{valoresatuais}
\begin{center}
\begin{tabular}{ccc}
\hline
\hline
Parameter&Mass hierarchy&True value\\ 
\hline
$\sin^2\theta_{12}$ &NH or IH&0.306\\
$\sin^2\theta_{13}$ &NH&0.02166\\
$\sin^2\theta_{13}$ &IH&0.02179\\
$\sin^2\theta_{23}$ &NH&0.441\\
$\sin^2\theta_{23}$ &IH&0.587\\
$\delta m^2_{21}$ &NH or IH&$7.5\times10^{-5}~\textrm{eV}^2$\\
$\delta m^2_{31}$ &NH&$2.524\times10^{-3}~\textrm{eV}^2$\\
$\delta m^2_{13}$ &IH&$2.514\times10^{-3}~\textrm{eV}^2$\\
$\delta_{\rm cp}$ &NH&$261\pi/180$\\
$\delta_{\rm cp}$ &IH&$277\pi/180$\\
\hline 
\hline
\end{tabular}
\end{center}
\end{table}

\begin{figure}[htbp]
\centering  
\includegraphics[width=0.9\textwidth]{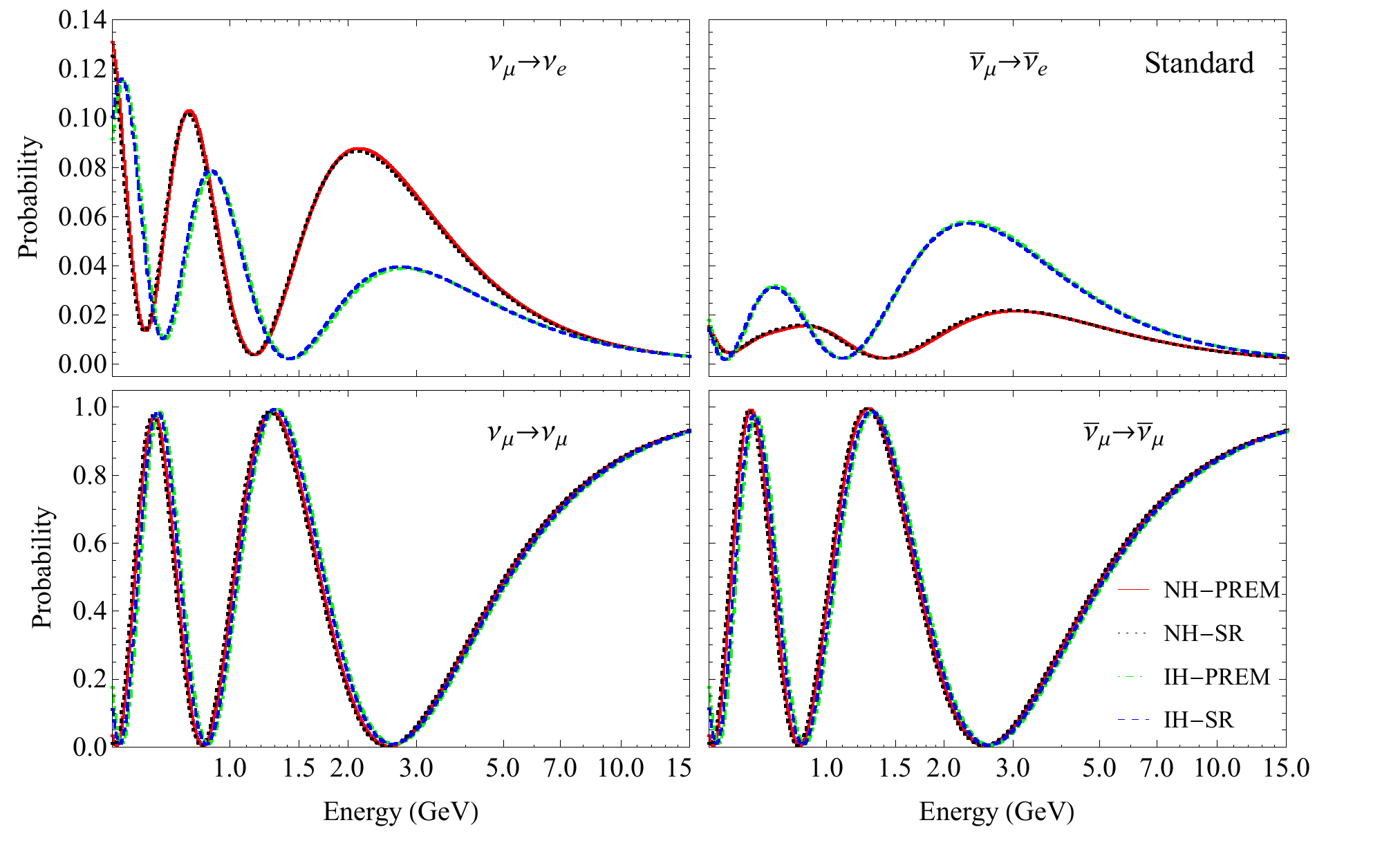}
\caption{Probability of oscillation for appearance and disappearance channels. We compare two different matter density profiles according to figure~\ref{fig:profile} for the possible mass hierarchies. See colors online.}~\label{fig:prob}

\includegraphics[width=0.9\textwidth]{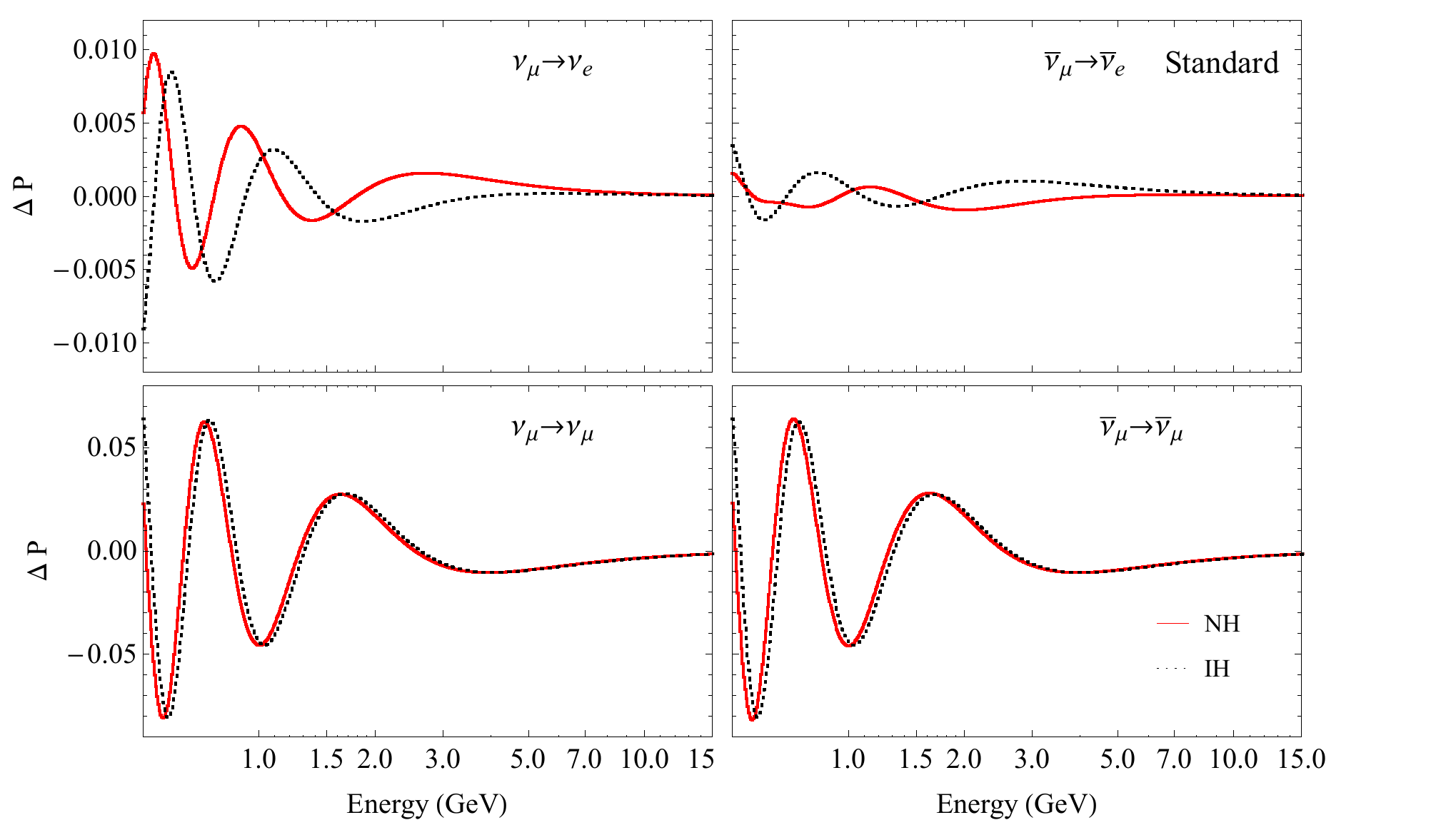}
\caption{Probability difference for appearance and disappearance oscillation channels. We calculate the probability difference of two different matter density profiles (see figure~\ref{fig:profile}) for the possible mass hierarchies. See colors online.}~\label{fig:deltaprob}
\end{figure}

In figure~\ref{fig:prob}, we show both the $\nu_{e}$ appearance and $\nu_{\mu}$ disappearance channel probability with the two different density profiles in the case of standard oscillation. There is no major difference in the oscillation probability due to different profile for the standard oscillation, both the normal (NH) and inverted hierarchy (IH). To understand it better, the differences in the probability for both channels are shown in figure~\ref{fig:deltaprob}.
In the appearance channel, where the oscillation probability is always less than 0.14, the difference is less than $\sim0.01$. In the disappearance channel, where the probability can basically reach the unity, $|\Delta P| < 0.09\,$. In both cases the relative difference is less than 10\%. This is
discussed in details in~\cite{Roe:2017zdw,Kelly:2018kmb}. However, the study of the matter profile effect can be more realistic if we investigate the observed number of events at the DUNE far detector. Most challenging and important question will be whether the detector is sensitive enough to measure a difference in the number of events.\\

Furthermore, we studied the effect of different density profile for the neutrino interaction beyond the standard model, known as non-standard interaction (NSI). Such framework is described in details in section~\ref{sec:nsi}. 

\section{Neutrino oscillation probability in matter with NSI} \label{sec:nsi}
Non-standard neutrino interaction refers to the interaction of neutrinos with the matter fermions. We consider effects that can be  phenomenologically described by neutral current (NC) type neutrino NSI of the form     
\begin{equation}
\label{nsilag}
{\cal L}_{NSI} = -2 \sqrt 2 G_F \epsilon_{\alpha \beta}^{f\, C} ~ [\bar \nu_\alpha \gamma^\mu P_L \nu_\beta] ~[\bar f \gamma_\mu P_C f]~,
\end{equation}
where $G_F$ is the Fermi constant, $\nu_{\alpha},\nu_{\beta}$ are neutrinos of different flavors. $\epsilon_{\alpha \beta}^{f\, C}$ is the NSI coupling with the neutrinos. The chiral projection operators are given by $P_{L} = (1 - \gamma_{5})/2$, $P_{C} = (1 \pm\gamma_{5})/2$.\\

In presence of NSI, the propagation of neutrinos is governed by a \schd-type equation with the effective Hamiltonian  
\begin{equation} 
{\mathcal H} = {\mathcal H}_{\mathrm vac} +  {\mathcal H}_{SI} +  {\mathcal H}_{NSI} 
\end{equation}

where ${\mathcal H}^{}_{\mathrm{vac}} $ is the vacuum Hamiltonian and
${\mathcal H}^{}_{\mathrm{SI}}, {\mathcal H}^{}_{\mathrm{NSI}}$ are
the effective Hamiltonians in presence of 
{SI alone and NSI} respectively. Thus the Hamiltonian can be written as,

\begin{equation}
 \label{hexpand} {\mathcal
H}^{}_{\mathrm{}} = \frac{1}{2E} \left\{ { U} \left(
\begin{array}{ccc}
0   &  &  \\  &  \delta m^2_{21} &   \\ 
 &  & \delta m^2_{31} \\
\end{array} 
\right) {U}^\dagger + 
 2E{V_{CC}}   \left(
\begin{array}{ccc}
1+ \epsilon_{ee}  & \epsilon_{e \mu}  & \epsilon_{e \tau}  \\ {\epsilon_{e\mu} }^ \star & \epsilon_{\mu \mu} &   \epsilon_{\mu \tau} \\ 
{\epsilon_{e \tau}}^\star & {\epsilon_{\mu \tau}}^\star & \epsilon_{\tau \tau}\\
\end{array} 
\right) \right\}  
 \end{equation}
where $\epsilon_{\alpha\beta}=|\epsilon_{\alpha\beta}|\exp(i\phi_{\alpha\beta})$ are the NSI parameters. Equation~\ref{hexpand} shows that the impact of NSI on the oscillation largely depends on the earth matter density profile.

\subsection{Impact of matter density profile on probability and NSI} \label{sec:impact}
Let us try to understand the effect of different matter density with NSI on the appearance channel, which is very crucial for DUNE experiment. The approximate expression for the oscillation probability for $\numu \to \nue$ considering NSI can be obtained by neglecting the higher order terms. In the case of a long baseline experiment like DUNE, one can safely neglect the smaller mass squared difference $\delta m^2_{21}$ in comparison to $\delta m^2_{31}$ since $\delta m^2_{21} L/(4 E) = \Delta_{21}L/2 \ll 1$ for a large range of values of $L$ and $E$ (especially above a GeV).  This ``one mass scale dominant'' (OMSD) approximation allows for a relatively simple exact analytic computation of the probability (as a function of only three parameters $\theta_{23},\theta_{13}$ and $\delta m^2_{31}$). In order to systematically take into account the effect of small parameters, the perturbation theory approach is used.\\

The probability for $\nu_{\mu} \to \nu_{e}$ channel is given by
\begin{eqnarray}
 P_{e\mu}^{{{NSI} }} &\simeq& 
 4 s_{13}^2 s_{23}^2 \, \left[\dfrac{ \sin^2 {(1-r_A )\Delta_{31} L/2}{}}{ (1-r_A)^2} \right] \nonumber\\
 && + \, 8  s_{13} s_{23} c_{23}
( |\epsilon_{e \mu}| c_{23} c_{\chi} - |\epsilon_{e \tau}| s_{23} c_{\omega} ) \, r_A \,\left[\dfrac{\sin {r_A \Delta_{31} L/2}}{r_A} ~
  \dfrac{ \sin {(1 - r_A) \Delta_{31} L/2}}{(1-r_A)} ~\cos \frac{ \Delta_{31} L}{2}
 \right] \nonumber\\
  &&
+ \, 8  s_{13} s_{23} c_{23}
( |\epsilon_{e \mu}| c_{23} s_{\chi} - |\epsilon_{e \tau}| s_{23} s_{\omega} )r_A \,\left[\dfrac{\sin {r_A \Delta_{31} L/2}}{r_A} ~
  \dfrac{\sin {(1 - r_A) \Delta_{31} L/2}}{(1-r_A) } ~\sin \frac{ \Delta_{31} L}{2}
 \right]\nonumber\\
 && + \, 8  s_{13} s_{23}^2 
( |\epsilon_{e \mu}| s_{23} c_{\chi} + |\epsilon_{e \tau}| c_{23} c_{\omega} )r_A \,\left[
 \dfrac{\sin^2 {(1 - r_A) \Delta_{31} L/2} }{(1-r_A)^2} 
 \right] \,,
 \label{pem}
 \end{eqnarray}
where $s_{ij} = \sin{\theta_{ij}}, c_{ij} = \cos{\theta_{ij}}$, $\Delta_{31} = \frac{\delta{m^{2}}_{31}}{2E}$, and $r_{A}= \frac{2EV_{CC}}{\delta{m^{2}}_{31}}$. Also, $c_\xi \,\, (s_{\xi}) = \cos\xi \,\,(\sin\xi)$ ($\xi=\chi,w$), $\chi = \phi_{e\mu} + \delta_{\rm cc}$, and $w = \phi_{e\tau} + \delta_{\rm cc}$. Only two NSI parameters ($\epsilon_{e\mu}$, $\epsilon_{e\tau}$) enter in this leading order expression which implies that the rest of the NSI parameters are expected to play a sub-dominant role. Equation (\ref{pem}) allows us to illustrate the qualitative impact of the NSI parameters on the appearance probability due to the different density profile.\\ 

The probability for $\nu_{\mu} \to \nu_{\mu}$ channel is given by
\begin{eqnarray}
\label{pmm}
 P_{\mu\mu}^{{{NSI}}}  &\simeq& 1 - s^2_{2\times {23}} \left[ \sin^2 \frac{\Delta_{31} L}{2} \right] \nonumber\\
 && - ~
 |\epsilon_{\mu\tau}| \cos \phi_{\mu\tau} s_{2 \times {23}} \left[ s^2 _{2 \times {23}} (r_A \Delta_{31} L) \sin {{\Delta_{31} L}{}} + 4  c^2_{2 \times {23}}  r_A \sin^2 \frac{ \Delta_{31} L}{2} 
 \right]\nonumber\\
 && +~ (|\epsilon _{\mu\mu}| - |\epsilon _{\tau\tau}|) s^2_{2 \times {23}} c_{2 \times {23}}\left[  \dfrac{r_A \Delta_{31} L}{2} \sin  {\Delta_{31} L}{} - 2  r_A \sin^2 \frac{\Delta_{31} L}{2} \right] \,,
 \end{eqnarray}
where $s_{2 \times {23}} \equiv \sin 2 \theta_{23} $ and $c_{2 \times {23}} \equiv \cos 2 \theta_{23} $. Note that the NSI parameters involving the electron sector do not enter this channel and the survival probability depends only on the three parameters $\epsilon_{\mu\mu}$, $\epsilon_{\mu\tau}$, and $\epsilon_{\tau\tau}$.\\

Let us now discuss the two limiting cases of the matter density profile. First: when $r_{A}\Rightarrow 0$, we recover the vacuum oscillation probability. Second: when $r_{A}\Rightarrow 1$, we are near resonance.\\

Inspired by the NSI discovery potential of DUNE discussed in~\cite{deGouvea:2015ndi} we analyze the differences in matter profile effect among four special cases of NSI that are listed in table~\ref{tab:nsinew}. We choose these particular cases because, except for Case 4, they are allowed by present data but at DUNE they would be a clear evidence of non-standard interactions. In Case 1, it is assumed that only off-diagonal epsilons and the phases containing $e$ ($\phi_{e\beta}$) are relevant. In Case 2, it is assumed that the new physics is in the diagonal epsilons ($\varepsilon_{ee}$ and $\varepsilon_{\tau\tau}$). In Case 3, $\varepsilon_{\alpha\mu}=0$ and the new physics would be in the electron and tau sector, either diagonal or off-diagonal. In Case 4, we investigate $\varepsilon_{\mu\tau}$ and its phase together with $\varepsilon_{ee}$, assuming $|\varepsilon_{\mu\tau}|=0.2$ which is in the limit of sensitivity for DUNE, although excluded by other experiments.
\begin{table}[htb]
\caption{Values of the NSI parameters for 4 different cases of interest. See text for details.} \label{tab:nsinew}
\begin{center}
\begin{tabular}{ccccccccc}
\hline
\hline
&$\varepsilon_{ee}$&$\varepsilon_{\tau\tau}$&$|\varepsilon_{e\mu}|$&$|\varepsilon_{e\tau}|$&$|\varepsilon_{\mu\tau}|$&$\phi_{e\mu}$&$\phi_{e\tau}$&$\phi_{\mu\tau}$\\ 
\hline
Case 1 & 0 & 0 & 0.15 & 0.3 & 0.05 & $\pi/3$ & $-\pi/4$ & 0 \\
Case 2 & -1.0 & 0.3 & 0 & 0 & 0 & 0 & 0 & 0 \\
Case 3 & 0.5 & -0.3 & 0 & 0.5 & 0 & 0 & $\pi/3$ & 0 \\
Case 4 & 0.5 & 0 & 0 & 0 & 0.2 & 0 & 0 & $-\pi/2$ \\
\hline 
\hline
\end{tabular}
\end{center}
\end{table}

\begin{figure}[htbp]
\centering
\includegraphics[scale=0.8]{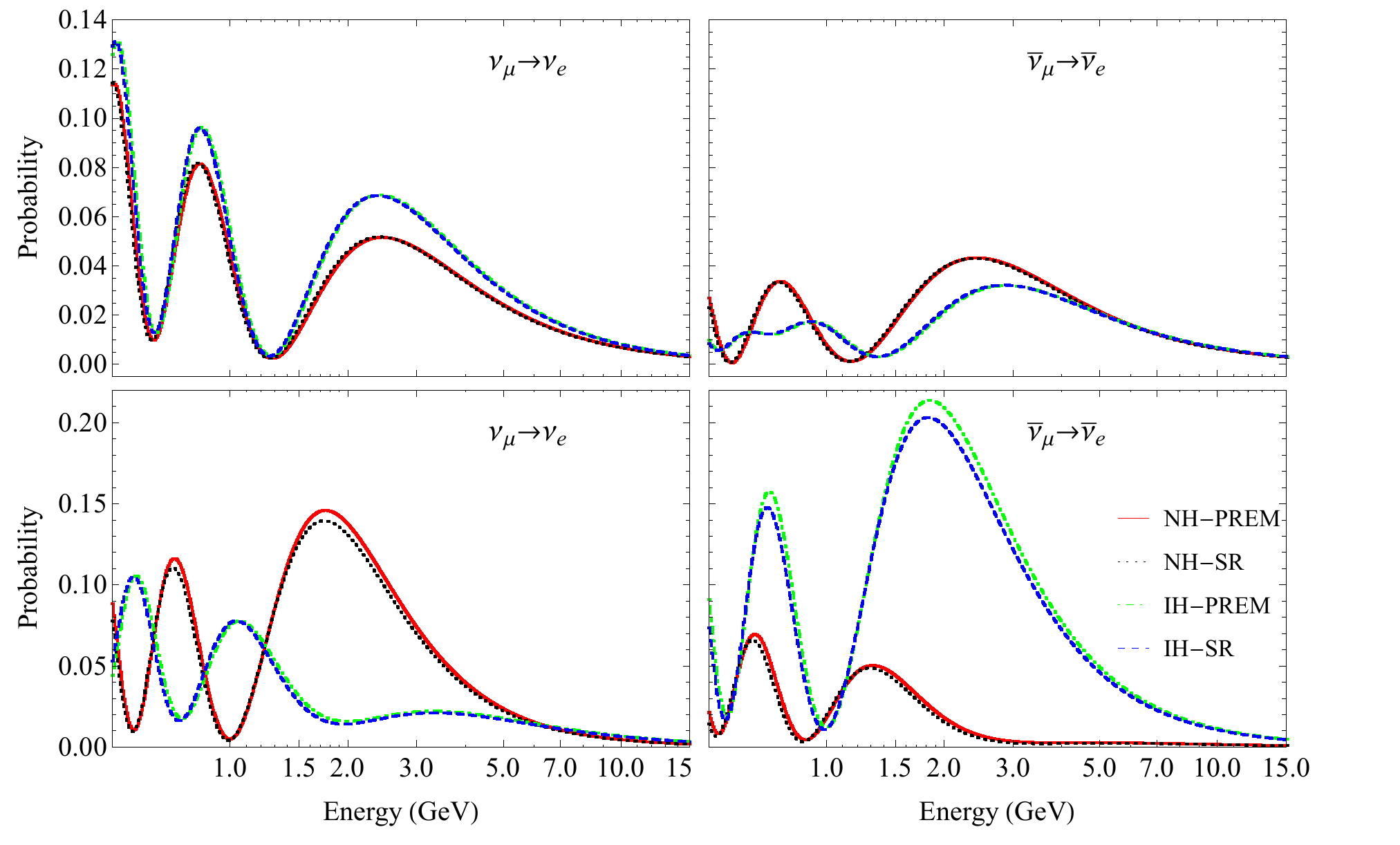}
\caption{Effect of NSI on the oscillation probability. In Case 2 (top), with diagonal parameters $\varepsilon_{ee}$ and $\varepsilon_{\tau\tau}$, we do not observe significant difference between PREM and SR profiles. In Case 3 (bottom), there is an interplay between electron and tau non-standard interactions so $\varepsilon_{ee, \tau\tau, e\tau}\neq0$ and $\varepsilon_{e\tau}$ has a complex phase of $\pi/3$. There is a visible difference in the probability for PREM and SR profiles. } \label{fig:prob23}
\end{figure}

\begin{figure}[htbp]
\centering  
\includegraphics[scale=0.8]{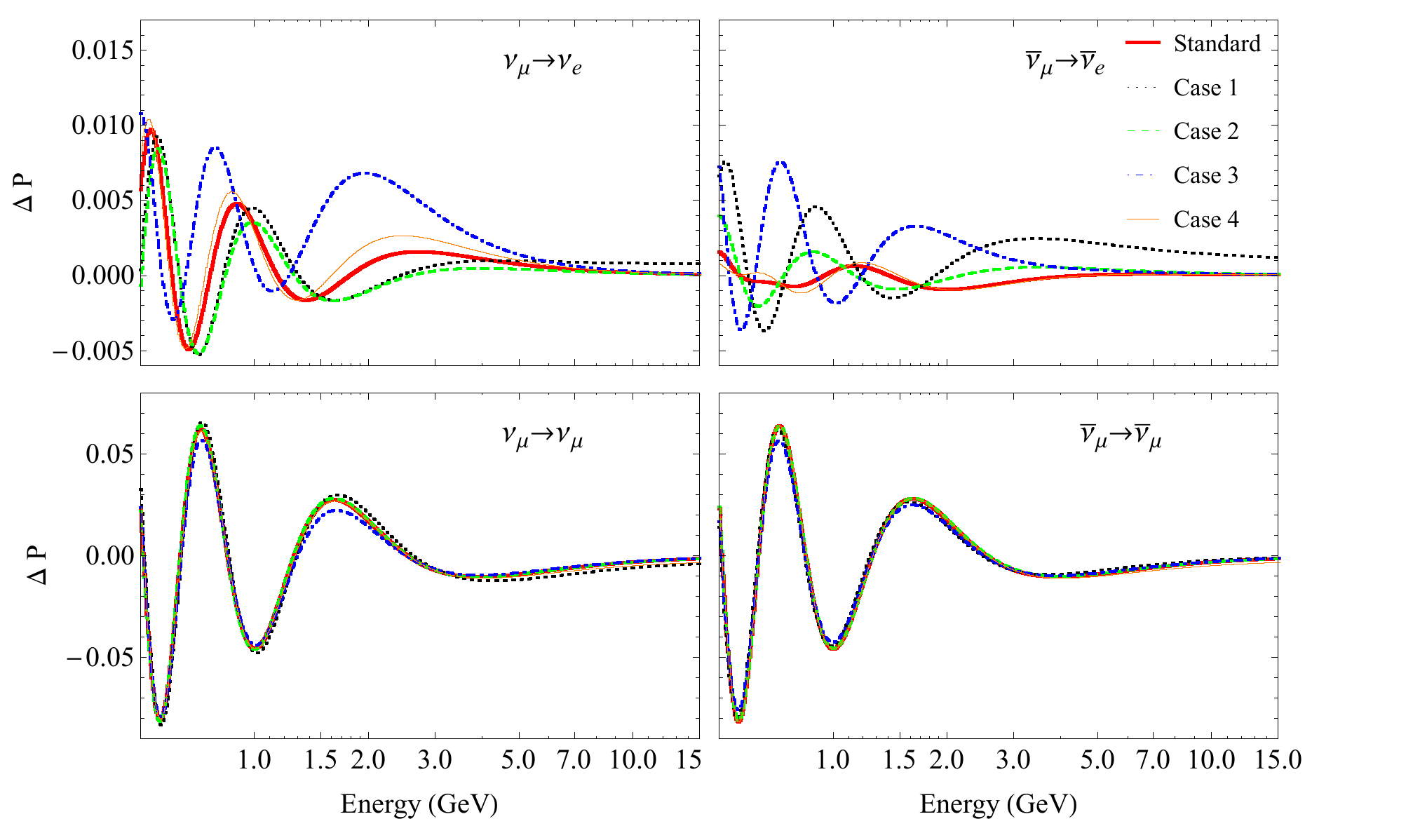}
\caption{Difference in the probability of oscillation due to propagation through different density profiles. We show the four cases for normal hierarchy. See the values considered in each case in table~\ref{tab:nsinew}.} \label{fig:deltaprobnsi}
\end{figure}

In figure \ref{fig:prob23}, we show the probability of oscillation for the PREM and the SR matter profiles in the appearance channels, for normal and inverted hierarchy. Among the four cases and oscillation channels of our study we choose to show these two cases and appearance channels because, as we see in figure~\ref{fig:deltaprobnsi}, Case 2 (3) presents the smaller (bigger) absolute probability difference $|\Delta P|$ in the energy range that is more relevant for the neutrino detection, around $2-3$~GeV.

In figure \ref{fig:deltaprobnsi}, we have shown the difference in the appearance (top) and disappearance (bottom) probability of oscillation between PREM and SR profile for four difference cases. The maximum absolute difference of the appearance channel occurs in Case 3 and Case 1. This nature can be easily understood from the equation (\ref{pem}). The leading order term of the off-diagonal NSI parameters, $\epsilon_{e\mu}$ and $\epsilon_{e\tau}$, have the maximum effect on the appearance probability. For Case 2 and Case 4, the zero values of $\epsilon_{e\mu}$ and $\epsilon_{e\tau}$ do not show any difference between the two profiles.  The difference in the disappearance channel is not significant.

There are not much differences on the appearance and disappearance probability for the standard oscillation (SI) with the different earth density profiles, PREM and SR, but there will be a significant effect on the oscillation probability if there is NSI.

\section{Event rates at DUNE far detector}
\label{sec:events}

In the simulation of neutrino propagation through matter we use the configuration for the DUNE CDR~\cite{Alion:2016uaj}. The expected event spectra are simulated using GLoBES~\cite{Huber:2004ka,Huber:2007ji}. We summarize the configurations below. The time of operation is 3.5 years for each neutrino and anti-neutrino beam with 1.2~MW of power and 0.56 efficiency. The mass of the detector is 40~kton. The processes and associated uncertainties are mentioned in table~\ref{tabelasistematicos}.
\begin{table}[htb]
\caption{All the channels considered to produce events in the detector. The associated uncertainty to each process is the same for neutrinos or anti-neutrinos.} \label{tabelasistematicos}
\begin{center}
\begin{tabular}{cccc}
\hline
\hline
Uncertainty&Channel&Interaction&Comment\\ 
\hline
2\% & $\nu_\mu \to \nu_e$ &CC& Signal, oscillation to $\nu_e$\\
5\% & $\nu_e \to \nu_e$ &CC& Background, beam contamination by $\nu_e$ \\
5\% & $\nu_\mu \to \nu_\mu$ &CC& Background, no oscillation to $\nu_e$ \\
20\% & $\nu_\mu \to \nu_\tau$ &CC& Background, oscillation to $\nu_\tau$ \\
10\% & $\nu_\mu/\nu_e \to X$ &NC& Background, interaction via NC\\
\hline
5\% & $\nu_\mu \to \nu_\mu$ &CC& Signal, no oscillation to $\nu_e$ \\
20\% & $\nu_\mu \to \nu_\tau$ &CC& Background, oscillation to $\nu_\tau$ \\
10\% & $\nu_\mu/\nu_e \to X$ &NC& Background, interaction via NC\\
\hline 
\hline
\end{tabular}
\end{center}
\end{table}

We use GLoBES with the Monte Carlo Utility Based Experiment Simulator (MonteCUBES) C library~\cite{Blennow:2009pk}, a plugin that replaces the deterministic GLoBES minimizer by a Markov Chain Monte Carlo method that is able to handle higher dimensional parameter spaces. We reproduce the analysis conducted in~\cite{Coloma:2015kiu} and~\cite{deGouvea:2015ndi} considering all the NSI parameters non negligible.\\

In the figures that follow we can understand what is the best way to quantitatively measure the difference between two models of density profile in case of NSI.\\

\begin{figure}[htbp]
\centering
\includegraphics[scale=0.8]{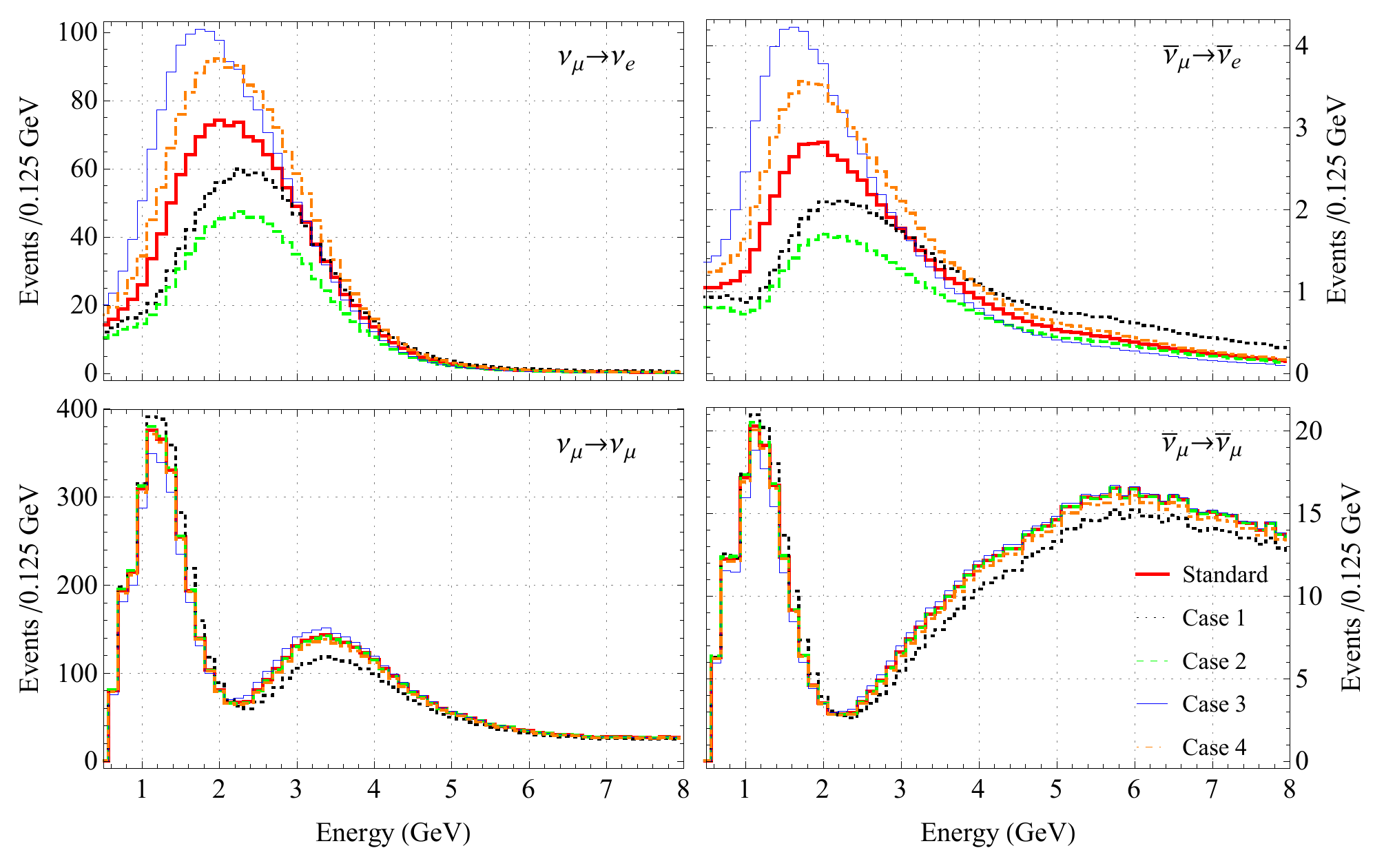}
\caption{Event rate with PREM profile and NH for different nsi cases. Top and bottom panels show $\nu_{e}$, ($\bar{\nu_e}$ on right) and  $\nu_{\mu}$ ($\bar{\nu_{\mu}}$ on right) events respectively. The effect of case 3 is maximum.}\label{fig:evtnhprem}
\includegraphics[scale=0.8]{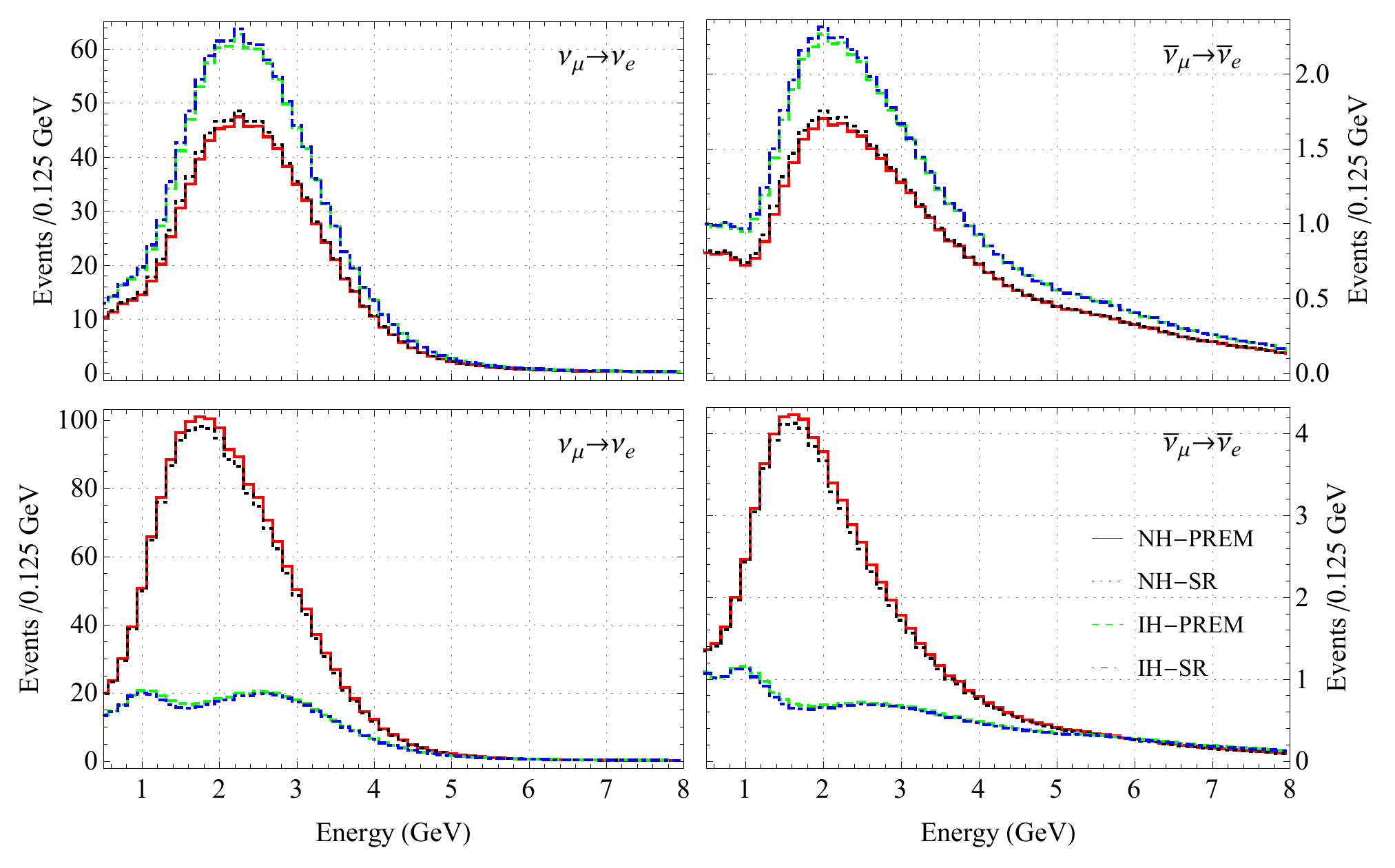}
\caption{Number of events per 0.125 GeV bin of energy. Case 2 (top) and 3 (bottom).
See table~\ref{tab:nsinew} and text for details on each NSI particular case.} \label{fig:nevents23}
\end{figure}

In figure~\ref{fig:evtnhprem}, we show the number of events with the PREM profile and NH. The event rates are shown with standard oscillation parameters along with the four different cases as mentioned earlier. It is clear from the figure, that the effect of Case 3 is maximum, as evidenced in the appearance probability plot. The effect of the NSI parameters are visible both on $\nu_{e}$ and $\nu_{\mu}$ events. 

In figure \ref{fig:nevents23}, we show the number of events per energy bin (0.125 GeV) and compare PREM versus SR profiles as well as normal versus inverted hierarchy. Electron neutrino events are shown for Case 2 (top) and for Case 3 (bottom) respectively. Maximum variation is observed in case of IH for Case 2, while NH has the most significant effect for Case 3. The same aspect is visible for both density profiles. We notice from figure~\ref{fig:evtnhprem} that these cases are interesting because they give the minimum and maximum event rates in the appearance channels. Later, we show that these cases result in the most relevant absolute event number difference $|\Delta N|$ due to the matter profile difference. \\

The difference in the number of events is
\begin{equation}
\Delta N = N_{\rm PREM} - N_{\rm SR} \,.
\end{equation}
The number of events has uncertainty defined as $\sigma_{\rm PREM}$ or $\sigma_{\rm SR}$ from
\begin{equation}
N_{\rm model} = \langle N_{\rm model}\rangle\pm\sigma_{\rm model}\,,
\end{equation}
where ``model'' is related to PREM or SR.\\

\begin{figure}[htbp]
\centering
\includegraphics[scale=0.8]{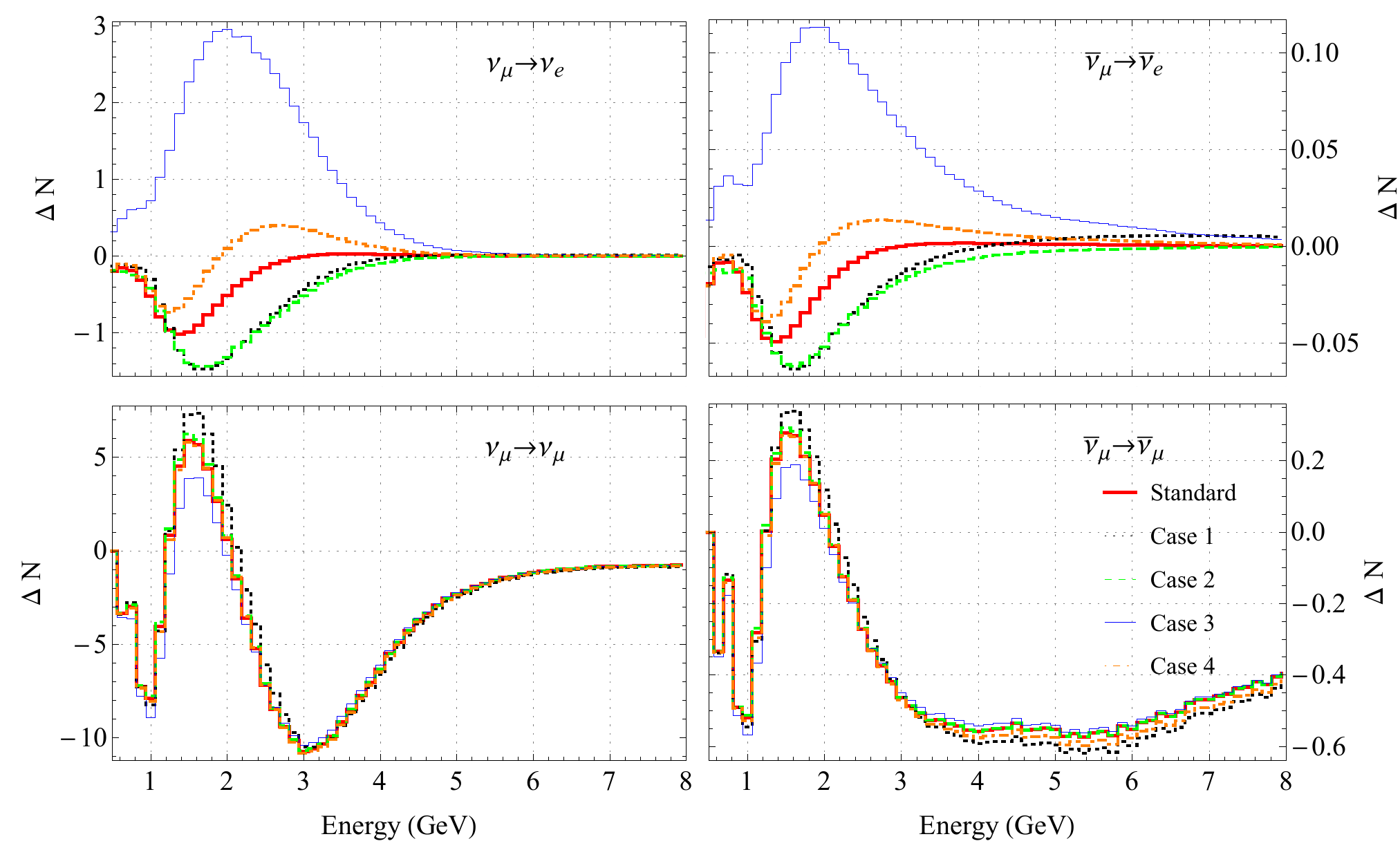}
\caption{Difference in the number of events per 0.125 GeV bin of energy, and for normal hierarchy. We calculate the number of events for PREM and SR density profiles. See table~\ref{tab:nsinew} and text for details on each NSI particular case. } \label{fig:deltan}
\includegraphics[scale=0.8]{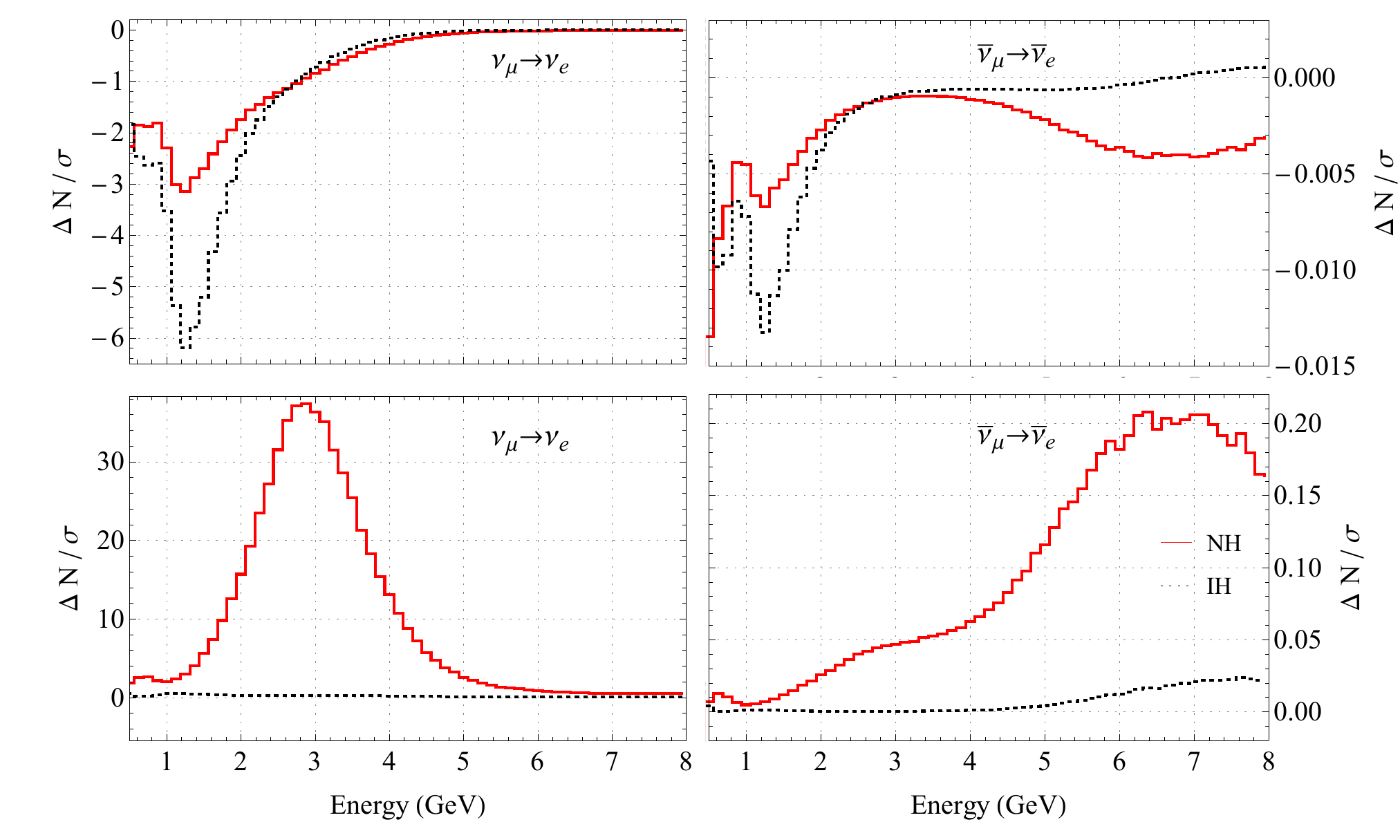}
\caption{Difference in the number of events with respect to the uncertainty ($\frac{\Delta N}{\sigma}$) per 0.125 GeV bin of energy. We show the Cases~2 (top) and 3~(bottom) results for normal and inverted hierarchy. See text for details.} \label{fig:dnsigma23}
\end{figure}

The difference in the number of events for the appearance and disappearance spectra is shown in figure~\ref{fig:deltan} for NH and the four different NSI instances. The difference of the number of events for the appearance spectra is maximum in Case 3 as discussed in the probability plot.\\ 

Besides the number of events and the difference in this number due to the different profiles of the matter crossed by the neutrinos, it is crucial to determine if the experiment is sensitive to this difference. We can decide this sensitivity calculating the ratio of $\Delta N$ and $\sigma$, where
\begin{equation}
\sigma = \sqrt{\sigma_{\rm PREM}^2+\sigma_{\rm SR}^2} \,.
\end{equation}
This is what we show in figure~\ref{fig:dnsigma23}. In Case~2, $|\Delta N|$ can be as big as $6\sigma$ for inverted hierarchy in the $\nu_e$ appearance channel. The most astounding though is that for Case~3 $\Delta N$ reaches around $30-40~\sigma$ for normal hierarchy in the appearance channel and energies of $2-3$~GeV.
We also checked, even though we do not show the plots here, that in both cases, 2 and 3, $|\Delta N|\approx4\sigma$ for both hierarchies in the $\nu_\mu$ disappearance channel, for energies around 2.5~GeV. 
\\

\begin{figure}[htbp]
\centering
\includegraphics[scale=0.8]{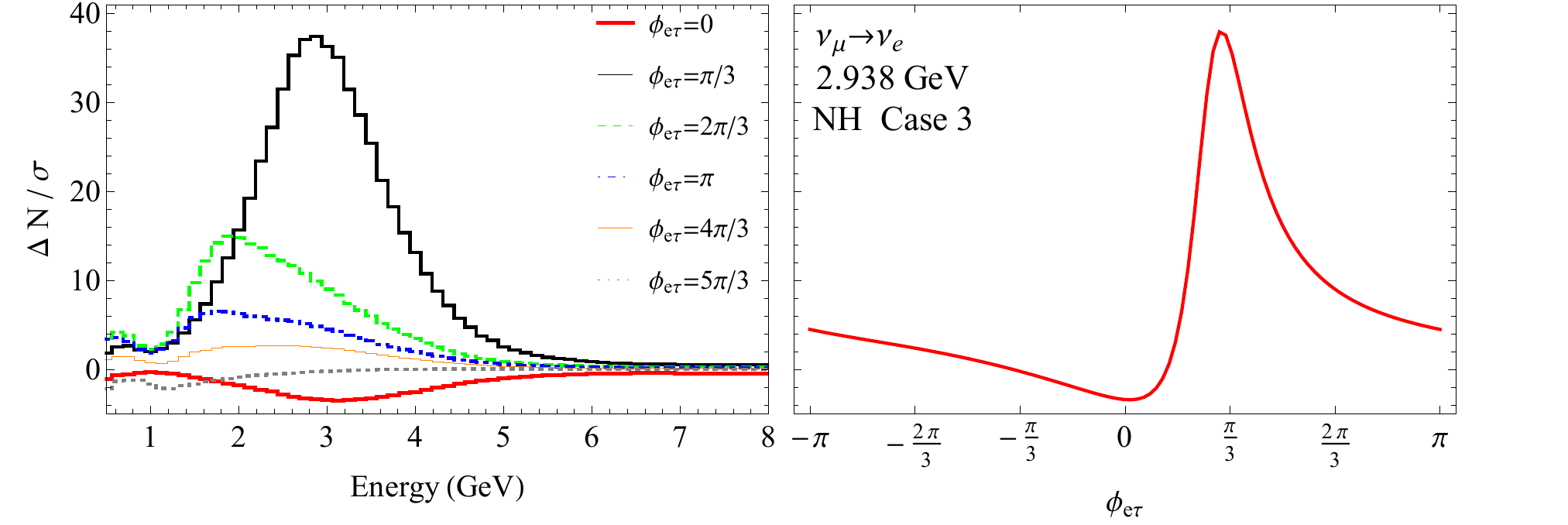}
\caption{Difference in the number of events with respect to the uncertainty ($\frac{\Delta N}{\sigma}$) per 0.125 GeV bin of energy. We calculate the number of events for PREM and SR density profiles, and for normal and inverted hierarchy. See table~\ref{tab:nsinew} and text for details on each NSI particular case.} \label{fig:phases}
\end{figure}

In figure \ref{fig:phases}, one sees how, considering Case~3, $\Delta N/\sigma$ depends on the value of the non-standard phase $\phi_{e\tau}$. It is interesting to notice that $\Delta N\approx0$ if $\phi_{e\tau}\approx-\frac{1}{3}\pi$ but $\Delta N$ is maximum if $\phi_{e\tau}\approx\frac{1}{3}\pi$, where $\frac{\Delta N}{\sigma}\approx38$. In this case, it would be extremely important to consider matter density profile with high precision.

\section{Discussion and conclusion}
\label{sec:conclude}

In this work we study the effects of different matter density profile for the standard and non-standard interactions in the measurement of neutrino oscillation at DUNE. First, we calculated the differences in oscillation probabilities at DUNE for both appearance and disappearance channels for two different density profiles in case of standard oscillation. There are not many significant differences observed between different profiles, both for the appearance and disappearance channels. Then, we investigate the effect with non-standard interactions.\\

We first derive the analytical expression of both the appearance and disappearance probabilities along with the matter density and NSI parameters. The off-diagonal NSI parameters have maximum effect on the appearance channel at the leading order level. It is also clear from the expressions in section~\ref{sec:impact} that any variation of matter density will have significant effect on the understanding or discovery of NSI.
Next, we show the effect of different matter density with two different sets of values of NSI parameters. The maximum difference of the oscillation is observed in the case where off-diagonal elements are present. The major difference is observed in Case 3 (see table~\ref{tab:nsinew}) due to the effect of $\epsilon_{e\mu}$ and $\epsilon_{e\tau}$.\\ 

From previous works~\cite{Roe:2017zdw,Kelly:2018kmb} we know that the baseline and average matter density used in the calculation is important to predict the number of neutrino events with accuracy. We consider carefully the effects of the matter density profile in the ability to calculate the number of events at DUNE with accuracy and how important it is to consider the profile precisely. We show that there is one combination of non-standard interaction parameters that results in a considerable difference in the number of events and consequently, if this happens to be the case in nature, which is Case 3 in table~\ref{tab:nsinew}, this will strongly affect the determination of the number of electron neutrino appearance events. The interesting feature is that this has a high dependence on the non-standard phase $\phi_{e\tau}$ and the effect would be observable preferably if this phase is approximately $\pi/3$.\\

In conclusion, we explored the effect of different matter density profiles for the understanding of NSI using the DUNE experiment. Our study clearly shows that, at least in one particular case, two different matter profiles will provide different event number results for this same set of NSI parameters. This degeneracy needs to be resolved with the clear understanding of the matter profile as well as the discovery of the hierarchy before calling any discovery of non-standard neutrino interactions.

\section*{Acknowledgements} 
The work of C.Moura was partially supported by Funda\c{c}\~ao de Amparo \`a Pesquisa do Estado de S\~ao Paulo (FAPESP), under the Grant No. 2014/19164-6. We also thank M.Guzzo, O.Peres, and D.Forero for important discussions. AC and JY are supported by the U.S. Department of Energy, HEP Award DE-SC0011686.

\bibliographystyle{ieeetr}
\bibliography{referencesnsi}

\end{document}